\documentstyle[pra,aps,manuscript]{revtex}
\input{epsf}

\begin{document}
\title{Realization of a Decoherence-free, Optimally Distinguishable Mesoscopic
Quantum Superposition.}
\author{Francesco De Martini, Fabio Sciarrino, and Veronica Secondi}
\address{Dipartimento di Fisica and Istituto Nazionale per la Fisica della Materia,\\
Universit\'{a} ''La Sapienza'', Roma 00185, Italy}
\maketitle

\begin{abstract}
We report the realization of an {\it entangled} quantum superposition of \ $%
M\sim 12$ photons by a high gain, quantum-injected optical parametric
amplification. The system is found {\em s}o highly resilient against
decoherence to exhibit directly accessible mesoscopic interference effects
at normal{\em \ }temperature. By modern tomographic methods the
non-separability and the quantum superposition are demonstrated for the
overall mesoscopic output state of \ the dynamic ''closed system''. The
device realizes the condition conceived by Erwin Schroedinger with his 1935
paradigmatic ''Cat'' apologue, a fundamental landmark in quantum mechanics.
\end{abstract}


\narrowtext

Since the golden years of quantum mechanics the interference of classically
distinguishable quantum states, first epitomized by the famous ''{\it %
Schroedinger-Cat}'' Apologue \cite{Schr35} has been the object of extensive
theoretical studies and recognized as a major conceptual paradigm of physics 
\cite{Cald83}. In modern times the sciences of quantum information (QI)\ and
quantum computation deal precisely with collective processes involving a
multiplicity of interfering states, generally mutually entangled and rapidly
de-phased by decoherence \cite{Zure81}. In many respects the implementation
of this intriguing classical-quantum condition represents today an unsolved
problem in spite of recent successful studies carried out mostly with atoms
and ions \cite{Brun96,Monr96}. The present work reports on a virtually
decoherence-free scheme based on the quantum-injected optical parametric
amplification (QI-OPA) of a single photon in a quantum superposition state
of polarization $(\pi )$, i.e. a $\pi -encoded$ qubit{\it \ }\cite
{DeMa98,DeMa01}. Conceptually, the method consists of transferring the well
accessible condition of quantum superposition of a single photon qubit, $N=1$%
, to a mesoscopic, i.e., multi-photon amplified state $M>>1$, here referred
to as a ''mesoscopic\ qubit'' (M-qubit). This can be done by injecting in
the QI-OPA the 1-photon qubit{\it , }$\alpha \left| H\right\rangle +\beta
\left| V\right\rangle $, here expressed in terms of two orthogonal $\pi -$%
states, e.g. horizontal and vertical linear $\pi $: $\left| H\right\rangle $%
, $\left| V\right\rangle $. In virtue of the general\ {\it information
preserving}\ property of the OPA,\ the generated state is found to be
entangled and to keep the {\it same} superposition character and the
interfering properties of the injected qubit \cite{DeMa98}. Since the
present scheme realizes the deterministic $1\rightarrow M$ {\it universal} 
{\it optimal quantum} {\it cloning} {\it machine} (UOQCM), i.e. able to copy 
{\it optimally} any unknown input qubit into $M>>1$ copies with the same
''fidelity'', the output state will be necessarily affected by {\it %
squeezed-vacuum noise }arising from the input vacuum field.

Let's refer to the apparatus: Fig. 1. The active element was a nonlinear
(NL)\ crystal slab (BBO:\ $\beta -$barium borate), 1.5 mm thick cut for Type
II\ phase-matching, able to generate by spontaneous parametric down
conversion (SPDC) $\overrightarrow{\pi }$-entangled photons pairs. The OPA 
{\it intrinsic phase} was set as to generate by SPDC {\it singlet} states on
the output modes, a condition assuring the {\it universality} of the present
cloning transformation \cite{DeMa98,DeMa02,Lama02,Pell03}.The excitation
source was a Ti:Sa Coherent MIRA mode-locked laser further amplified by a
Ti-Sa regenerative REGA\ device ({\bf A})\ operating with pulse duration $%
180fs$ at a repetition rate $250kHz$, average power $1W$. By ({\bf A}) the
OPA nonlinear (NL)\ ''gain'' $g$\ was enhanced by a factor $\simeq 17$
respect to earlier experiments \cite{DeMa02,Lama02,Pell03}. The output beam,
frequency doubled by second harmonic generation (SHG) provided the
excitation beam with UV\ wavelength (wl) $\lambda _{P}=397.5nm$ and energy
per pulse $E_{UV}^{HG}=1\mu J$. The ''seed'' photons pairs were emitted,
with a coherence time $\approx 500fs$, by a OPA process acting towards the
right hand side (r.h.s.)\ of Fig.1 with equal wl's $\lambda =795nm$ over two
spatial modes $-{\bf k}_{1}$ and $-{\bf k}_{2}$ owing to a SPDC process
excited by the UV\ beam associated with mode $-{\bf k}_{p}$ with wl $\lambda
_{p}$. The UV\ beam was back-reflected over the mode ${\bf k}_{p}\ $onto the
NL\ crystal by a spherical mirror ${\bf M}_{p}$, with $\mu $-metrically
adjustable position {\bf Z}, thus exciting the main OPA ``cloning''\ process
towards the left hand side of Fig.1. By the combined effect of two
adjustable optical UV wave-plates (wp's) $(\lambda /2$ + $\lambda /4)$
acting on the projections of the linear polarization ${\bf \pi }_{p}^{UV}$
on the optical axis of the BBO crystal for the $-{\bf k}_{p}$ and ${\bf k}%
_{p}$ excitation processes, the ''seed'' SPDC\ excitation was always kept at
a low level while driving the main OPA to a large gain $(HG)$ regime.
Precisely, by smartly unbalancing the orientation of the axes of the UV $%
wp^{^{\prime }}s$, the SPDC emission probability towards the r.h.s. of Fig.1
of 2 simultaneous photon pairs was always kept below the one of single pair
emission by a factor $\sim 3\times 10^{-2}$. One of the photons of the
''seed'' SPDC\ pair, back-reflected by a fixed mirror ${\bf M}$, was {\it %
re-injected} after a $\overrightarrow{\pi }$-flipping by a $\lambda /4$ wp,
onto the NL crystal over the input mode ${\bf k}_{1}$, while the other
photon emitted over mode $(-{\bf k}_{2})\ $excited the detector $D_{T}$, the 
{\it trigger }of the overall {\it conditional }experiment. The entangled
state of the ''seed'' pair after $M$-reflection and $\overrightarrow{\pi }$%
-flipping was: $\left| \Phi ^{-}\right\rangle _{-k2,k1}$= $2^{-1/2}\left(
\left| H\right\rangle _{-k2}\left| H\right\rangle _{k1}-\left|
V\right\rangle _{-k2}\left| V\right\rangle _{k1}\right) $. In virtue of the
nonlocal correlation acting on the ''seed'' modes $-{\bf k}_{1}$ and $-{\bf k%
}_{2}$, the input qubit was prepared on mode ${\bf k}_{1}$ in the {\it pure }%
state $\left| \Psi \right\rangle _{in}$= $\alpha \left| H\right\rangle
_{k1}+\beta \left| V\right\rangle _{k1}$, $\left| \alpha \right| ^{2}+\left|
\beta \right| ^{2}=1$ by the combined action of the $\lambda /2$ wp, of $%
\lambda /4$ wp $(WP_{T})$, of the adjustable {\it Babinet Compensator} $(B)$
and of a polarizing beam-splitter ($PBS_{T}$) acting on mode $-{\bf k}_{2}$.
This device allowed all orthogonal transformations $\widehat{U}_{X}$, $%
\widehat{U}_{Y}$, $\widehat{U}_{Z}$ on the Bloch sphere of the input qubit:\
Fig.1, inset. The Quantum State Tomography (QST)\ devices $T_{T},T_{i}(i$=$%
1,2)$ were equipped with equal single-photon fiber coupled SPCM-AQR14-FC
detectors $(D)\;$and equal interference filters with bandwidth $\Delta
\lambda =4.5nm$ were placed in front of each $D$: Fig.1-inset.

Let's re-write $\left| \Psi \right\rangle _{in}$ in terms of Fock product
states: $\left| H\right\rangle _{k1}$= $\left| 1\right\rangle _{1H}\left|
0\right\rangle _{1V}\left| 0\right\rangle _{2H}\left| 0\right\rangle
_{2V}\equiv \left| 1,0,0,0\right\rangle $; $\left| V\right\rangle _{k1}$= $%
\left| 0,1,0,0\right\rangle $, accounting for $1$ photon on the input ${\bf k%
}_{1}$ with different orthogonal $\overrightarrow{\pi }^{\prime }s$ and
vacuum on the input ${\bf k}_{2}$. It evolves into the output state $\left|
\Psi \right\rangle =\widehat{U}\left| \Psi \right\rangle _{in}$ according to
the main OPA\ unitary $\widehat{U}$ process \cite{Pell03}$.$ The output
state $\widetilde{\rho }=(\left| \Psi \right\rangle \left\langle \Psi
\right| )$ over the modes ${\bf k}_{1}$, ${\bf k}_{2}$ of the QI-OPA
apparatus is found to be expressed by the {\it M-qubit}: 
\begin{equation}
\left| \Psi \right\rangle =\alpha \left| \Psi \right\rangle ^{H}+\beta
\left| \Psi \right\rangle ^{V}  \label{Outputstate}
\end{equation}
where: $\left| \Psi \right\rangle ^{H}$= $\sum\limits_{i,j=0}^{\infty
}\gamma _{ij}\sqrt{i+1}\left| i+1,j,j,i\right\rangle $, $\gamma _{ij}\equiv
\cosh ^{-3}g(-\Gamma )^{i}\Gamma ^{j}$, $\Gamma \equiv \tanh g$, $\left|
\Psi \right\rangle ^{V}$= $\sum\limits_{i,j=0}^{\infty }\gamma _{ij}\sqrt{j+1%
}\left| i,j+1,j,i\right\rangle $, being $g$\ the NL\ gain{\em \ }\cite
{DeMa02}. These interfering {\it entangled}, multi-particle states are {\it %
ortho-normal}, i.e. $\left| ^{i}\left\langle \Psi \mid \Psi \right\rangle
^{j}\right| ^{2}$= $\delta _{ij}$ $\left\{ i,j=H,V\right\} $\ and {\it pure, 
}i.e. fully represented by the operators: $\rho ^{H}=(\left| \Psi
\right\rangle \left\langle \Psi \right| )^{H}$, $\rho ^{V}=(\left| \Psi
\right\rangle \left\langle \Psi \right| )^{V}$.\ Hence the {\it pure} state $%
\left| \Psi \right\rangle $ is a quantum superposition of two multi-photon 
{\it pure} {\it states }and bears the {\it same} superposition properties of
the injected qubit. In addition, it is highly significant in the present
context to consider the output {\it pure} state of the {\it overall}
apparatus, including the ''trigger'' that enters in the dynamics through the
Bell state $\left| \Phi ^{-}\right\rangle _{-k2,k1}$. The state $\left|
\Sigma \right\rangle $, commonly referred to as the ''{\it Schroedinger Cat
State}'' \cite{Schleich01}, expresses the entanglement of all output modes $%
{\bf k}_{1},$ ${\bf k}_{2}$ and $-{\bf k}_{2}$, thus eliciting a peculiar
cause-effect dynamics within the overall ''closed'' system: 
\begin{equation}
\left| \Sigma \right\rangle \equiv 2^{-1/2}\left( \left| H\right\rangle
_{-k2}\left| \Psi \right\rangle ^{H}-\left| V\right\rangle _{-k2}\left| \Psi
\right\rangle ^{V}\right)
\end{equation}

\bigskip In the pictorial context of the Cat apologue, $\left| \Sigma
\right\rangle $ expresses the correlations established within the ''closed
box'' between the ''microscopic'' state of the decaying particle that
triggers the release of \ the deadly poison and the ''macroscopic'' state of
the threatened animal \cite{Schr35,Cald83}.\ The {\it entanglement entropy }$%
{\cal E}(\left| \Sigma \right\rangle )$ of {\it \ }$\widetilde{\rho }%
^{\prime }\equiv $ $\left| \Sigma \right\rangle \left\langle \Sigma \right| $
is expressed by the Von Neuman entropy of either the $-k_{2}$ or $OPA$
subsystem: ${\cal E}(\left| \Sigma \right\rangle )$= $S(\widetilde{\rho }%
_{-k2})$= $S(\widetilde{\rho }_{opa})$=$1$, being $\widetilde{\rho }_{-k2}$= 
$Tr_{opa}(\widetilde{\rho }^{\prime })$, $\widetilde{\rho }_{opa}$= $%
Tr_{-k2}(\widetilde{\rho }^{\prime })$ and $S(\widetilde{\rho }_{j})$= $%
-Tr\left( \widetilde{\rho }_{j}\log _{2}\widetilde{\rho }_{j}\right) $ \cite
{Benn96,Pell03}. The maximal attainable value is ${\cal E}(\left| \Sigma
\right\rangle )=1$ for the output bipartite system.

The experimental investigation of the multiphoton superposition and
entanglement implied by Eqs 1, 2 was carried out by means of the $T_{i},T_{T%
\text{ }}$ devices according to a ''loss method'' first applied to SPDC by 
\cite{Eise04}. The beams associated with the output modes ${\bf k}_{i}(i$=$%
1,2)$ were highly attenuated to the {\it single-photon} level by the two low
transmittivity BS's $(At)$ in Fig.1. Since the bipartite entanglement
affecting the multiphoton ${\bf k}_{i}$, $(i$=$1,2)$ also implies a
correlation of the single photons detected on either modes and since it is
impossible to create or enhance entanglement by {\it local} operations, e.g.
by the loss mechanism acting over each mode, the entanglement detected at
the {\it single-photon} level over ${\bf k}_{i}$, $(i$=$1,2)$ necessarily
implies the same property to affect the same modes in the {\it multi-photon}
condition. In agreement with the ''loss method'' we investigated by two
different QST\ experiments the {\it reduced} density matrices $\rho $ and $%
\rho ^{\prime }$, i.e. the {\it single-photon} counterparts of the pure,
multi-photon states $\widetilde{\rho }$ , $\widetilde{\rho }^{\prime }$
given by Eqs.1, 2. The\ experimental results, reported in Figs. 2, 3
respectively, are compared with the corresponding theoretical $\rho ^{th}$
and $\rho ^{\prime th}$ which have been calculated by a numerical algorithm
performing the multiple tracing of $\widetilde{\rho }$, $\widetilde{\rho }%
^{\prime }$ over all photons discarded by the $At$ devices on the modes $%
{\bf k}_{i}(i=1,2)$. Theoretical details on this most useful algorithm and
on the overall experiment will be given in a forthcoming comprehensive paper 
\cite{Camin05}. In order to carry out the calculations properly, the maximum
value of the ''gain'' $g_{\exp }$ and of the overall quantum efficiencies $%
\eta _{i}$ of the detection apparatuses acting on ${\bf k}_{i}(i=1,2)$ were
measured. It was found: $g_{\exp }=1.19\pm 0.05$ and $\eta _{1}$=$(4.9\pm
0.2)\%$; $\eta _{2}$=$(4.2\pm 0.2)\%$.

Let us now address the main goal of the present work, i.e. the detection and
characterization of the output states. The interference (IF)\ character of
the output field implied by the quantum superposition character of the input
qubit $\left| \Psi \right\rangle _{in}=2^{-1/2}\left( \left| H\right\rangle
+e^{i\varphi }\left| V\right\rangle \right) $ was detected simultaneously in
the basis $\left| \pm \right\rangle \equiv 2^{-1/2}(\left| H\right\rangle
\pm \left| V\right\rangle )$ over the output ''cloning'' mode, ${\bf k}_{1}$
and ''anticloning'', ${\bf k}_{2}$ by the $2-D$ coincidences $[D_{i},D_{T}]$
(square marks in Fig 1-inset) and $[D_{i}^{\ast },D_{T}]$ (circle marks) $(i$%
=$1,2)$. Precisely, the IF\ fringe patterns shown in Fig.1 correspond to $%
\widehat{U}_{Z}$ transformations on the input Bloch sphere, i.e. implying
changes of the phase $\varphi $. The fringe ''visibility'' $({\cal V)}$
measured over ${\bf k}_{1}$ was found to be gain-dependent ${\cal V}%
_{1}^{th}(g)$=$(1+2\Gamma ^{2})^{-1}$as predicted by theory \cite{DeMa98}.
The experimental value ${\cal V}_{1}$=$(32\pm 1)\%$ should be compared with
the theoretical one: ${\cal V}_{1}^{th}(g_{\exp })$=$42\%$. By setting $g$=$%
\Gamma $=0 the effective visibility of the input qubit was${\cal \ }$%
measured: ${\cal V}_{in}\approx 87\%$. The ${\cal V-}$value for the ${\bf k}%
_{2}$ mode ${\cal V}_{2}$=$(13\pm 1)\%$ should be compared with the
theoretical:$\ {\cal V}_{2}^{th}=33\%$ \cite{DeMa98}. These discrepancies
are attributed to unavoidable walk-off effects in the NL slab spoiling the
critical superposition of the injection and pump pulses in the bi-refringent
active region. In the quantum-injected $HG$\ regime, the{\em \ }overall
average number of the stimulated emission photons\ {\it per pulse} over $%
{\bf k}_{i}(i$=$1,2)$ was found, $M=(11.1\pm 1.3)$, a result consistent with
the value of $g$ measured by an entirely different experiment. Precisely,
the average number of photons generated on the cloning mode was: $%
M_{C}=6.1\pm 0.9$ and the average ''fidelity'', obtained by the
corresponding V-value on the same mode, was: $F_{C}=(1+{\cal V}%
_{1})/2=66.2\pm 0.5$. Note that for $M\rightarrow \infty $, viz. $%
g\rightarrow \infty $ and $\Gamma \rightarrow 1$, the ''fringe visibility''
and the ''fidelity'' attain the asymptotic values ${\cal V}_{1}^{th}$= $%
{\cal V}_{2}^{th}$= $33\%$ and $F_{C}$= $F_{AC}$= $(2/3)$. In the present
experiment, in absence of reliable photon number-resolving detectors, the
measurement of $M$ and $M_{C}$ was transformed into a detection rate
measurement as $\eta _{1}\sim \eta _{2}\ll 1$.

A most insightful state analysis was provided by a full QST\ study of\ the
output $\rho $ and $\rho ^{\prime }$, as said. Figure 2{\bf \ }shows the QST
analysis of the {\it reduced }output{\it \ }state $\rho $ determined by the
set of input $\left| \Psi \right\rangle _{in}$: $\{\left| H\right\rangle $, $%
\left| V\right\rangle $, $\left| \pm \right\rangle \}$. The experimental
data $\rho ^{\exp }$ shown by Fig. 2-{\bf b,} were obtained by a 3-D
coincidence method $\left[ D_{T},D_{1},D_{2}\right] $ for different settings
of the QST\ setups $T_{i}$. The good agreement between theory and experiment
is expressed by the measured average Uhlmann ''fidelity'': ${\cal F}(\rho
^{\exp },\rho ^{th})\equiv \lbrack Tr(\sqrt{\rho ^{\exp }}\rho ^{th}\sqrt{%
\rho ^{\exp }})^{%
%TCIMACRO{\UNICODE[m]{0xbd}}%
%BeginExpansion
{\frac12}%
%EndExpansion
}]^{2}$=$(96.6\pm 1.2)\%$. In Fig. 2-{\bf a, b }the structure of the $%
4\times 4$ matrices $\rho $ shows the relevant quantum features of the
output state. For{\bf \ }instance, the highest peak on the diagonals
expressing the quantum superposition of the input state shifts from the
position $\left| \phi \phi ^{\bot }\right\rangle \left\langle \phi \phi
^{\bot }\right| $ to $\left| \phi ^{\bot }\phi \right\rangle \left\langle
\phi ^{\bot }\phi \right| $ in correspondence with the OPA excitation by any
set of orthogonal injection states $\{\left| \phi \right\rangle $, $\left|
\phi ^{\bot }\right\rangle \}$, i.e. represented by the maxima and minima of
the IF\ fringe patterns of Fig.1, inset. The experimental patterns of Fig. 2-%
{\bf b }obtained{\bf \ }by{\bf \ }injection of different basis sets $%
\{\left| H\right\rangle ,\left| V\right\rangle \}$ and $\{\left|
+\right\rangle ,\left| -\right\rangle \}$ indeed confirm the
''universality'' of this process, i.e. reproducing identically for any
couple of orthogonal input states. As expected,\ the IF{\em \ }fringe
pattern of Fig 1-inset was found to disappear in absence of the injection
qubit (rhombo marks in Fig.1-inset) \cite{Camin05}.\ The application to all $%
\rho $ matrices shown in Fig.2-{\bf b} of the Peres-Horodecki Positive
Partial Transpose $(PPT)$\ criterion ensures the non-separability of the
reduced state $\rho $ and then necessarily of the corresponding ''true'' 
{\it multiphoton} output state $\widetilde{\rho }$, as said \cite
{Pere96,Eise04}. Indeed the minimal eigenvalue of the transpose of the
''theoretical'' matrices $\rho ^{th}$ of Fig. 2-{\bf a }was found {\it %
negative}, $\lambda _{\min }$=$-0.046${\bf , }a result reproduced by{\bf \ }%
all{\bf \ }$\rho ^{\exp }$ reported in Fig. 2-{\bf b}.{\bf \ }For instance%
{\bf \ }for{\bf \ }$\rho ^{\exp }$ determined by $\left| \Psi \right\rangle
_{in}=\left| -\right\rangle $ it was found: $\lambda $=$-0.014\pm 0.0025$
for $g_{\exp }\sim 1.19$. Let's turn our attention to the state $\rho
^{\prime }$ correlating {\it all output} modes: Fig. 3-{\bf b}. The QST\
reconstruction was achieved by the devices $T_{T}$, $T_{i}$ again with a
large fidelity: ${\cal F}(\rho ^{\prime \exp },\rho ^{\prime th})$= $%
(85.0\pm 1.1)\%$. The quantum superposition\ of the output state is
expressed here by the off-diagonal elements of both matrices: $\rho ^{\prime
th}$, $\rho ^{\prime \exp }$. Once again, the nonseparability of $\rho
^{\prime }$ for the bipartite system $-{\bf k}_{2}$ and $\left\{ {\bf k}_{1},%
{\bf k}_{2}\right\} $ was proved by the PPT\ method, a sufficient criterion
for 3-qubit mixed states. Again, the minimal eigenvalue of the transpose of
both matrices $\rho ^{\prime }$ in Fig. 3 was found {\it negative}: $\lambda
_{\min }^{\prime }=-0.024$ and $\lambda _{\exp }^{\prime }=-0.021\pm 0.004$.
This proves the nonseparability of the ''true'' tri-partite pure state $%
\widetilde{\rho }^{\prime }$, Eq.2, correlating the trigger and the
multiphoton ${\bf k}_{i}$ modes within the ''closed box''.

A striking property of the present system is its extreme resilience to
de-coherence as shown by the interference patterns of Fig.1 \cite{Zure81}.\
Consequently, unlike other systems involving atoms or superconductors \cite
{Brun96,Monr96} the mesoscopic superposition is {\it directly accessible }at
the output of the apparatus at {\it normal}, i.e.''room'', temperature (T).
This lucky result is partially attributed to the {\it minimum}
Hilbert-Schmidt $(d)$ ''distance'' on the {\it phase-space }of the
interfering states realized here: $d(\rho ^{H};\rho ^{V})$ = $Tr\left[ (\rho
^{H}-\rho ^{V})^{2}\right] $= $2$ \cite{Nie00}. Since in our system the
de-coherence can only be determined by stray reflection losses on the single
output surface of the NL\ crystal, a number of photons in the range $%
(10^{2}\div 10^{3})$ could be easily excited in quantum superposition. The
limited, i.e. {\it optimal,} distinguishability of the mesoscopic states is
attributed to our {\it single particle, cloningwise }$\pi -$measurement
method. However, the {\it exact} distinguishability implied by the
orthogonality of $\left| \Psi \right\rangle ^{H}$ and $\left| \Psi
\right\rangle ^{V}$ could be possibly attained if any POVM identifying in a 
{\it cumulative fashion} all $M$\ particles{\em \ }involved in the
interference could be found.

In summary, we have\ demonstrated the quantum interference of {\it mesoscopic%
}, {\it orthonormal}, {\it pure} states in agreement with the original
Schroedinger's proposal \cite{Schr35} and with our quantum theoretical
results, Eqs.1, 2.{\em \ }In the near future, the adoption of
''periodic-poled'' nonlinearities is expected to further increase $g$ by a
factor $\geq 2$ and then the value of $M$ by at least an order of magnitude.
On a conceptual side our system is expected to open a new trend of studies
on the persistence of the validity of crucial laws of quantum mechanics for
entangled mixed-state systems of increasing complexity \cite{Brun96}, on the
realization of \ GHZ\ processes and on the violation of Bell inequalities in
the multi-particles regime \cite{Reid02}. In addition, our method may
suggest a plausible signal-amplification model for the establishment of
collective coherence effects in complex biological systems at normal T, e.g.
within the search of any non-computable physical process in the
self-conscious brain \cite{Pen94}. We thank Marco Caminati and Riccardo
Perris for experimental help within the tomographic reconstructions and
Serge Haroche, Peter Knight, Wojciech Zurek, Vlado Buzek and Wolfgang
Schleich for enlightening discussions. Work \ supported by the FET EU
Network on QI Communication (IST-2000-29681: ATESIT), INFM (PRA\ ''CLON'')\
and by MIUR (COFIN 2002).

\vskip8mm

\parindent=0pt

\parskip=3mm

Figure 1. Layout of the {\it quantum-injected OPA }apparatus; $T_{i}$, $%
(i=1,2)$ Quantum State Tomographic (QST)\ setups; Bloch sphere
representation of the input qubit. Inset: mesoscopic interference fringe
patterns measured over the modes ${\bf k}_{i}$ Vs the input phase $\varphi $
for the $\widehat{U}_{Z}$ map. The continuous lines express the best fit
results.

Figure 2. QST plots of the {\it reduced} output state over the 2 output
modes ${\bf k}_{i}$, $(i=1,2)$ conditioned by the injection of a
polarization qubit $\left| \Psi \right\rangle _{in}$ on mode ${\bf k}_{1}.$ (%
{\bf a}) Theoretical simulation. ({\bf b}) Experimental results. The
imaginary components are negligible in the given scale.

Figure 3. QST plots of the reduced overall output state $\rho ^{\prime th}$%
over the 3 modes $-{\bf k}_{2}$, ${\bf k}_{i}$, $(i=1,2)$ under the
condition of lossy channels for the multiphoton modes ${\bf k}_{i}$. The
experimental $\rho ^{\prime \exp }$ was reconstructed by measuring $64$
three-qubit observables and by applying a linear inversion reconstruction.
Each QST run lasted $60s$ and yielded a maximum $1866$ threefold counts for
the $\left| HHV\right\rangle $ projection. The uncertainties of the data
were evaluated by a numerical simulation assuming Poissonian fluctuations.
The imaginary plots are reported at the r.h.s. of the figure.

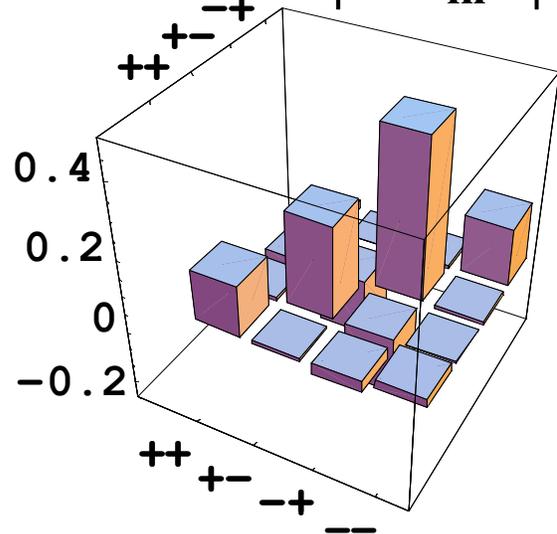
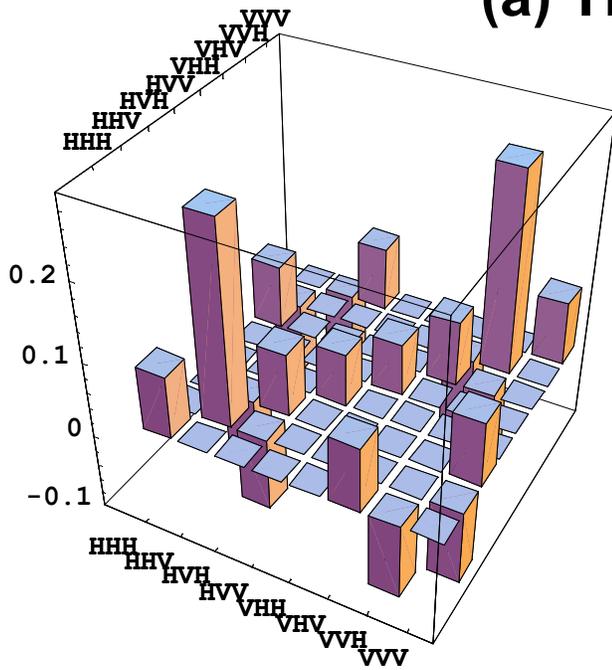
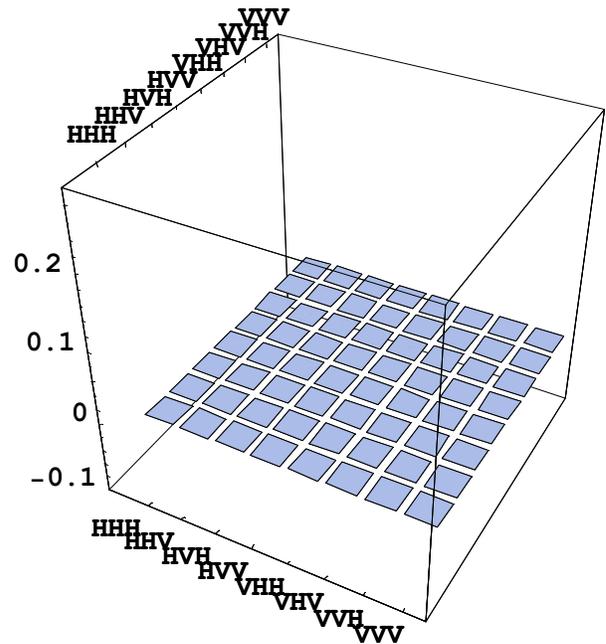
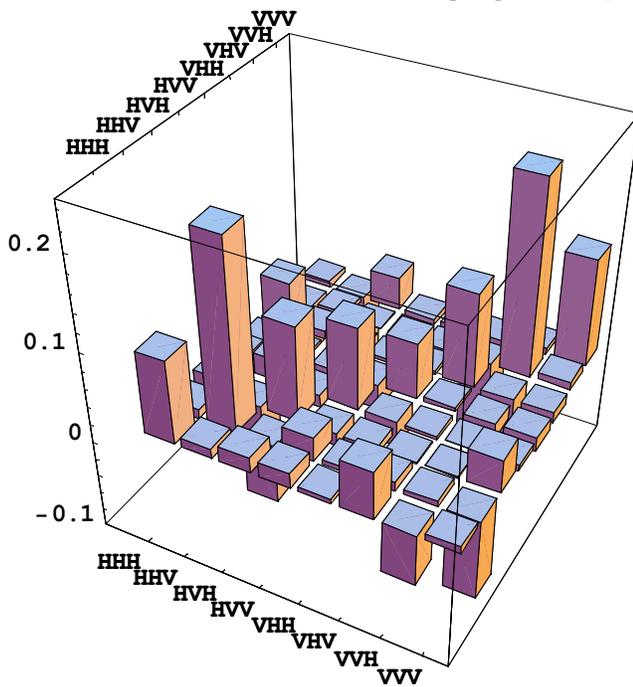
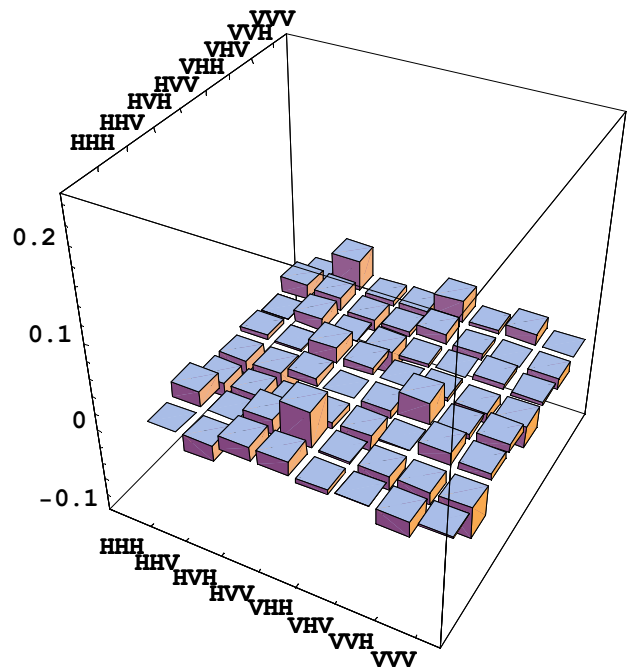
\end{document}